\documentclass[%
 aip,
 graphicx,
 amsmath,amssymb,
 reprint,%
]{revtex4-1}

\usepackage{graphicx}
\usepackage{dcolumn}
\usepackage{bm}

\usepackage[utf8]{inputenc}
\usepackage[T1]{fontenc}
\usepackage{mathptmx}
\usepackage{etoolbox}
\usepackage{color}
\usepackage{multirow}
\usepackage{ulem}
\usepackage{comment}

\usepackage[acronym, shortcuts]{glossaries}
\newacronym{BSG}{BSG}{borosilicate glass}
\newacronym{ASG}{ASG}{aluminosilicate glass}
\newacronym{CPT}{CPT}{coherent population trapping}
\newacronym{ALD}{ALD}{atomic layer deposition}
\newacronym{MEMS}{MEMS}{micro-electro-mechanical-systems}

\makeatletter
\def\@email#1#2{%
 \endgroup
 \patchcmd{\titleblock@produce}
  {\frontmatter@RRAPformat}
  {\frontmatter@RRAPformat{\produce@RRAP{*#1\href{mailto:#2}{#2}}}\frontmatter@RRAPformat}
  {}{}
}%
\makeatother
\begin{document}

\preprint{AIP/123-QED}

\title{Reduction of helium permeation in microfabricated cells using aluminosilicate glass substrates and Al$_2$O$_3$ coatings}

\author{C. Carl\'e}
\affiliation{FEMTO-ST Institute, Universit\'e de Franche-Comt\'e, CNRS, Besançon, France}
\author{S. Keshavarzi}
\affiliation{FEMTO-ST Institute, Universit\'e de Franche-Comt\'e, CNRS, Besançon, France}
\author{A. Mursa}
\affiliation{FEMTO-ST Institute, Universit\'e de Franche-Comt\'e, CNRS, Besançon, France}
\author{P. Karvinen}
\affiliation{Department of Physics and Mathematics, University of Eastern Finland, Joensuu, Finland}
\author{R. Chutani}
\affiliation{IEMN UMR 8520, Universit\'e Lille, CNRS, Centrale Lille, Universit\'e Polytechnique Hauts-de-France, Lille, France}
\author{S. Bargiel}
\affiliation{FEMTO-ST Institute, Universit\'e de Franche-Comt\'e, CNRS, Besançon, France}
\author{S. Queste}
\affiliation{FEMTO-ST Institute, Universit\'e de Franche-Comt\'e, CNRS, Besançon, France}
\author{R. Vicarini}
\affiliation{FEMTO-ST Institute, Universit\'e de Franche-Comt\'e, CNRS, Besançon, France}
\author{P. Abb\'e}
\affiliation{FEMTO-ST Institute, Universit\'e de Franche-Comt\'e, CNRS, Besançon, France}
\author{M. Abdel Hafiz}
\affiliation{FEMTO-ST Institute, Universit\'e de Franche-Comt\'e, CNRS, Besançon, France}
\author{V. Maurice}
\affiliation{IEMN UMR 8520, Universit\'e Lille, CNRS, Centrale Lille, Universit\'e Polytechnique Hauts-de-France, Lille, France}
\author{R.~Boudot}
\affiliation{FEMTO-ST Institute, Universit\'e de Franche-Comt\'e, CNRS, Besançon, France}
\author{N. Passilly}
\affiliation{FEMTO-ST Institute, Universit\'e de Franche-Comt\'e, CNRS, Besançon, France}
\email{nicolas.passilly@femto-st.fr}


\begin{abstract}
The stability and accuracy of atomic devices can be degraded by the evolution of their cell inner atmosphere. Hence, the undesired entrance or leakage of background or buffer gas, respectively, that can permeate through the cell walls, should be slowed down. In this work, we investigate helium permeation in microfabricated alkali vapor cells filled with He and whose windows are made of \ac{BSG} or \ac{ASG}. The permeation is then derived from routine measurements of the pressure-shifted hyperfine transition frequency of an atomic clock. We first confirm that \ac{ASG} reduces He permeation rate by more than two orders of magnitude, in comparison with \ac{BSG}. In addition, we demonstrate that Al$_2$O$_3$ thin-film coatings, known to avoid alkali consumption in vapor cells, can also significantly reduce He permeation. The permeation through \ac{BSG} is thereby reduced by a factor 110 whereas the one through \ac{ASG} is decreased by a factor up to 5.8 compared to uncoated substrates. These results may contribute to the development of miniaturized atomic clocks and sensors with improved long-term stability or sensitivity.
\end{abstract}

\maketitle
Alkali vapor cells are core elements of compact atomic devices such as clocks~\cite{Micalizio:Metrologia:2012, MAH:APL:2018}, magnetometers~\cite{Budker, Boto:Nature:2018}, gyroscopes~\cite{Jau:PRL:2017}, microwave electrometers~\cite{Sedlacek:Nature:2012}, or THz-imaging systems~\cite{Wade:Nature:2017} which can reach outstanding levels of accuracy and sensitivity. Vapor cells are also key components of a wide variety of quantum physics experiments, for the implementation of laser-cooling platforms~\cite{Lindquist:1992}, quantum teleportation demonstrations~\cite{Krauter:Nature:2013} or the achievement of spin-squeezed~\cite{Mitchell:PRL:2010} and entangled~\cite{Julsgaard:Nature:2001} states. The last twenty years have witnessed a renewed interest in vapor cells with the advent of their microfabricated version~\cite{Kitching:APL:2002, Liew:APL:2007, Douahi:EL:2007, Maurice:2017, Bopp:JPP:2021, Maurice:NMN:2022}. The latter, when combined with integrated photonics, has stimulated the development of low-power and easy-to-deploy chip-scale atomic devices~\cite{Kitching:APR:2018}, with real-world applications and tremendous prospects.

In an atomic cell device, it is critical to ensure that the probed atomic sample evolves in a well-controlled and stable environment. Variations of the cell inner pressure resulting either from background gas entrance or buffer gas leakage have to be avoided to prevent any perturbation of the probed atomic transition. In microwave atomic clocks, a buffer gas is typically added to extend the interrogation duration past the transit time. While adding this gas is beneficial in terms of short-term stability, it imparts a frequency shift proportional to the buffer gas pressure on the clock~\cite{Beverini:1981, Kozlova:PRA:2011} and imposes this pressure to be constant to ensure long-term stability~\cite{Camparo:2005, Camparo:2007, Abdullah:APL:2015}. In microcell optical frequency references~\cite{Hummon:Optica:2018, Newman:Optica:2018, Gusching:JOSAB:2021, Gusching:OL:2023}, the presence of contaminants in the cell induces a broadening of the optical resonance~\cite{Pitz:PRA:2009} that prevents benefiting from the thin natural linewidth of the transition. In trapped-ion or cold atom systems, degradation of the cell background pressure reduces the trapped ion or magneto-optical trap lifetime and jeopardizes the signal-to-noise ratio of the detected signal~\cite{Rushton:2014, McGilligan:2022}.

The cell atmosphere stability is then likely affected from gas permeation, especially of helium due to its low density, its volatility (smallest mono-atomic gas) and its significant natural concentration in the earth atmosphere (5.24~ppm)~\cite{Rogers:1954, Norton:1957, Altemose:1961, Rushton:2014}. For instance, the limitation of the long-term stability of an optical clock based on the Rb 778~nm two-photon transition has been attributed to He permeation~\cite{Lemke:2022}. In standard cold atom or trapped-ion systems, where ultra-high vacuum background pressure levels are required, preventing the science cell from being polluted by helium gas often requires using of an active external ion pump. This approach, suitable for the demonstration of laser cooling in lab-scale platforms, compact transportable systems~\cite{Garrido:2019} or even \ac{MEMS} cells~\cite{McGilligan:APL:2020}, is not compatible with the demonstration of a fully-miniaturized boardable atomic device. In this context, chip-scale ion pumps are being developed~\cite{Basu:2016}, but, along with increasing their complexity, the intrinsic generation of large magnetic field close to the atomic sample may limit the development of high-precision instruments.

Since passive getter pumps do not interact with helium because of its noble character~\cite{Scherer:2012, Boudot:SR:2020}, the main solution to prevent helium from entering the cell while keeping compact devices consists in constructing the cell or the science chamber with low-permeation materials. Several remarkable studies have recently been reported in this domain including a miniaturized vacuum package based on a metal-ceramic package and sapphire windows for a trapped ion atomic clock~\cite{Schwindt:2016}, a compact and passively pumped vacuum package combining titanium body and sapphire windows, sustaining cold atoms for more than 200 days~\cite{Little:2021} or the development of a stand-alone vacuum ceramic cell maintaining a cold atom cloud for more than one year~\cite{Burrow:2021}. Yet, the above-mentioned studies are all based on standard cells, and do not benefit from the collective fabrication and mass-production capabilities offered by \ac{MEMS} technologies. 

Microfabricated alkali vapor cells consist typically of a silicon-etched cavity sandwiched between two anodically-bonded glass wafers~\cite{Kitching:APL:2002, Liew:APL:2007, Douahi:EL:2007, Hasegawa:SA:2011, Maurice:2017, Abdullah:APL:2015, Vicarini:SA:2018, Newman:Optica:2018}. \ac{BSG} is well-adapted for anodic bonding to silicon and is widely used due to its wafer-form availability. However, if silicon is rather hermetic~\cite{Rushton:2014}, \ac{BSG} is known to exhibit a non-negligible gas permeation rate, for noble gases such as Ne~\cite{Abdullah:APL:2015, Carle:OE:2023}, and especially He~\cite{Kroemer:OE:2015, Dellis:2016}. Microfabricated cells with significantly reduced He permeability have then been reported where the cell windows were made of \ac{ASG} substrates~\cite{Dellis:2016, Newman:2021}. In Ref.~\citenum{Dellis:2016}, helium permeation rate was derived from the frequency drift of a \ac{CPT} clock, while the physics package with a cell filled with Ar and N$_2$ was placed inside a chamber filled with He. The positive frequency drift was then consistent with a progressive entrance of helium into the cell.

In the present study, we investigate helium permeation in microfabricated alkali vapor cells, with two types of glass substrates (\ac{BSG} or \ac{ASG}), through long-term measurements of the pressure-shifted hyperfine transition frequency of a \ac{CPT} atomic clock based on cells filled with He. Then, He permeation is monitored as a progressive decrease of the clock frequency along time, evidence of He going out of the cell. Moreover, we propose to add coatings onto the cell windows, as complementary permeation barriers. For this purpose, thin-film Al$_2$O$_3$ coatings, usually implemented in \ac{MEMS} cells to avoid alkali disappearance with time~\cite{Woetzel:2015, Karlen:2017}, are deposited onto the glass substrate prior to the bonding with the silicon preform. We show that Al$_2$O$_3$ layer can reduce He leak rates up to a factor of 110, when cell windows are made of \ac{BSG}. For cells using \ac{ASG}, which already exhibits He permeation rates about 450 times smaller than those obtained with \ac{BSG}, alumina coatings bring a further improvement by a factor of nearly 6 leading to a reduction of the He permeation rate by more than three orders of magnitude in comparison with standard cells made of bare \ac{BSG}.

Let us remind first the basic formalism used in this manuscript to study He permeation from microfabricated cells. The evolution of the buffer gas pressure $P(t)$ in the microcell can be described by an exponential decay, such as:
\begin{equation}
P (t) = P_{ext} - (P_{ext} - P_{in})e^{-t/\tau},
\label{eq:eq1}
\end{equation}
with $t$ the time, $P_{in}$ the initial buffer gas pressure in the cell at $t=0$~s,
 $P_{ext}=3.98$~mTorr the partial pressure of He in the atmosphere, and $\tau$ the time constant described as:
\begin{equation}
\tau=\frac{V~d}{K~A~P_{ref}},
\label{eq:eq2}
\end{equation}
where  $V$ is the volume of the cell cavities and $P_{ref}=$~760~Torr is the atmospheric reference pressure. Since, as mentioned above, permeation of the cell silicon body is negligible, we assume that He permeates only through the glass windows. Consequently, $A$ and $d$ are the surface and the thickness, respectively, of the cell windows. Finally, $K$ is the glass permeation rate.

The latter can then be derived from the change of He pressure in the cell which in turn induces an evolution of the clock transition frequency due to the temperature-dependent buffer gas pressure shift $\Delta \nu_{bg}$, given by:
\begin{equation}
\Delta\nu_{bg} = P[\beta + \delta(T-T_{0}) + \gamma(T-T_{0})^{2}],
\label{eq:eq3}
\end{equation}

with $P$ the buffer gas pressure (at a reference temperature $T_0$ of 0$^{\circ}$C), $\beta$ the pressure coefficient, $\delta$ the linear temperature coefficient, $\gamma$ the quadratic temperature coefficient and $T$ the cell temperature (set at 70$^{\circ}$C in the first part of this study). For Eq.~(\ref{eq:eq3}), we use He coefficients reported in Ref.~\citenum{Beverini:1981} ($\beta$~=~1185$\pm$46~Hz/Torr, $\delta$~=~1.49$\pm$0.15~Hz/Torr.K). Hence, the permeation rate $K$ through the cell windows made of different glasses and eventually coated with Al$_2$O$_3$ can be determined by deriving the buffer gas pressure along time from the monitoring of the frequency of an atomic clock using the different cell configurations.

\begin{figure}[t]
\centering
\includegraphics[width=0.99\linewidth]{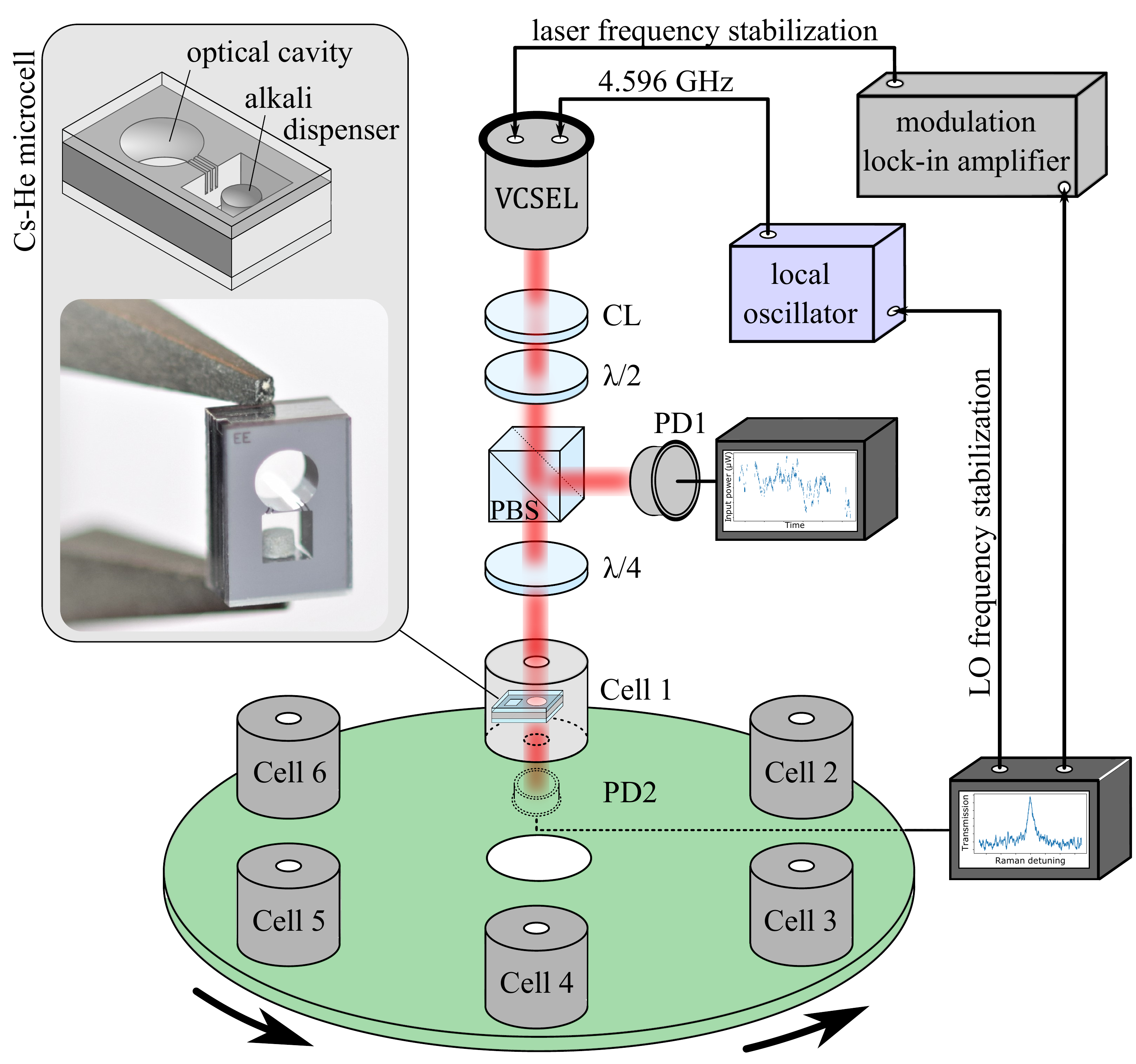}
\caption{(Color online) Multi-cell \ac{CPT} clock experimental setup. 6 cell physics packages are embedded into mu-metal shields and mounted onto a rotation platform allowing their sequential probing. CL: Collimation Lens, $\lambda$/2: Half-Wave Plate, $\lambda$/4: Quarter-Wave Plate, LO: Local Oscillator. PD1 and PD2 are photodiodes. The figure inset shows the cell architecture, based on a structured Si spacer sandwiched between two glass substrates.}
\label{fig:fig1}
\end{figure}

For this purpose, we implemented a table-top \ac{CPT} clock featuring an automated rotatable platform carrying 6 cell physics packages. Such a configuration allows investigating a significant number of cells in a regular manner during extended measurement campaigns. Figure~\ref{fig:fig1} shows a schematic of the multi-cells \ac{CPT} clock. It is based on a single and fixed vertical optical axis in front of which 6 different cell physics packages, mounted on a rotation platform, can be sequentially positioned. The laser source is a commercially-available vertical-cavity surface-emitting laser (VCSEL) tuned on the Cs D$_1$ line at $\lambda=~$895~nm~\cite{Kroemer:AO:2016}. The VCSEL is bonded onto a dedicated card that embeds a bias-tee through which a 4.596~GHz signal is injected to modulate the laser and produce \ac{CPT} optical sidebands. A lens and a quarter wave plate are used to collimate and circularly polarize the laser beam. The rotation platform is actuated by a stepper motor in order to switch periodically the cell under test. Each package consists of a \ac{MEMS} cell inserted in a duralumin piece temperature-stabilized at 70$^{\circ}$C, surrounded by a static magnetic field to raise the Zeeman degeneracy and protected by a mu-metal magnetic shielding. The laser light is detected at the output of the cell under test by a photodiode. The acquired signal is used in a first servo loop that acts on the laser dc current to stabilize the laser frequency to the center of the Doppler-broadened optical resonance. A second servo loop is used to lock the microwave interrogation signal frequency to the \ac{CPT} resonance. In this experiment, the 4.596 GHz signal is generated by a commercial microwave synthesizer, referenced to an active hydrogen maser used as a reference for frequency shifts and stability measurements.

The cell technology is comparable to the one described in Refs~\citenum{Hasegawa:SA:2011} and \citenum{Vicarini:SA:2018}.
Atom-light interaction takes place in a cylindrical cavity (shown in the inset of Fig.~\ref{fig:fig1}) with a diameter of $2000\pm20$~$\mu$m and a length of $1500\pm10$~$\mu$m. The side of the squared dispenser cavity is $1600\pm20$~$\mu$m long. The two glass substrates are $510\pm10$~$\mu$m thick and are made either of \ac{BSG} Borofloat\textsuperscript{\textregistered}33 from SCHOTT or of \ac{ASG} SD2 from HOYA. Each wafer typically gathers 215 cells, from which between 10 to 20 are filled with Cs dispensers. For this study, eight wafers of Cs-He cells have been fabricated in total.

\begin{figure}[t]
\centering
\includegraphics[width=0.99\linewidth]{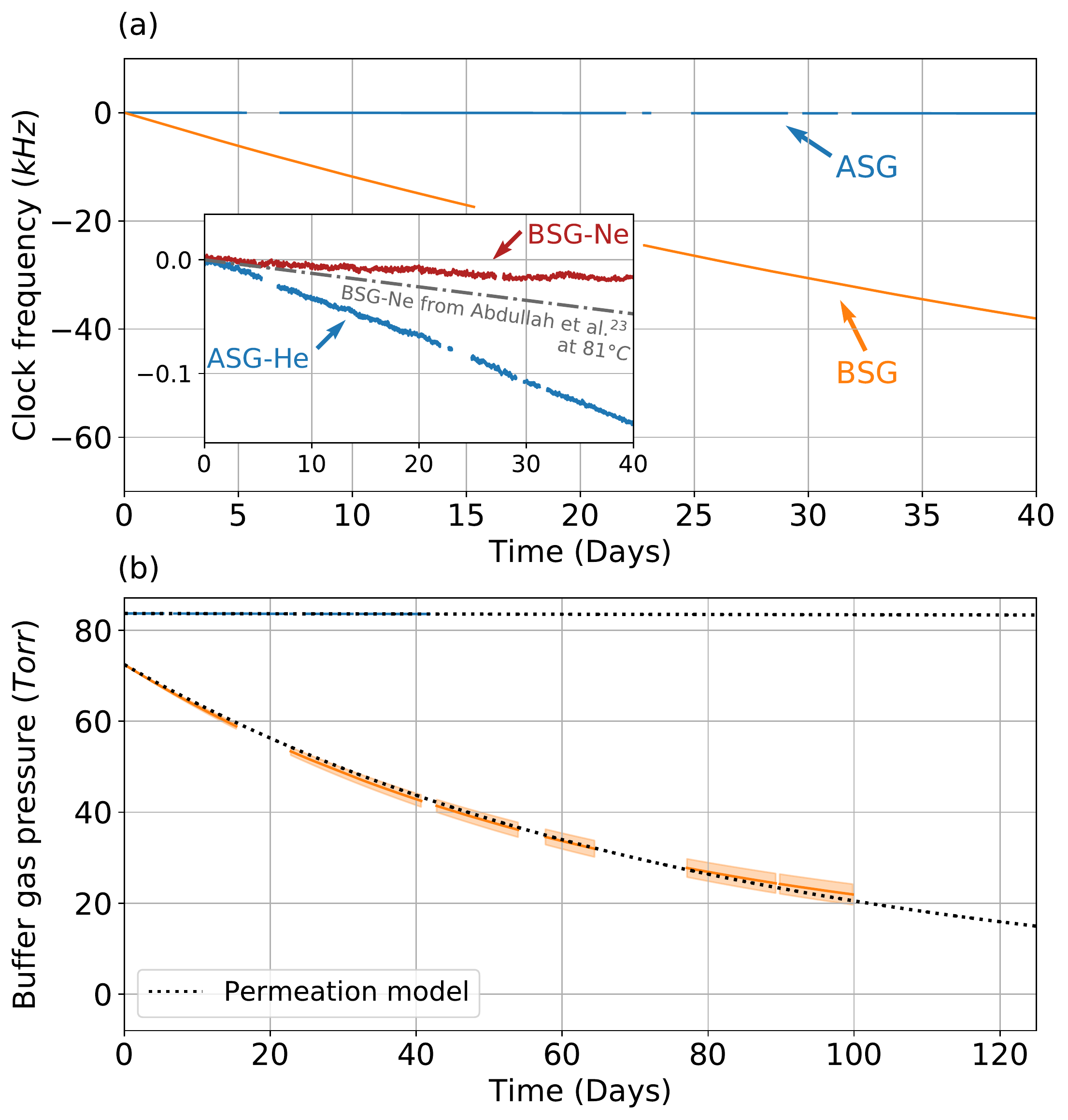}
\caption{(Color online) (a)~Temporal trace of the clock frequency using microfabricated Cs-He cells based on \ac{BSG} or \ac{ASG}. In the inset, a clock frequency measurement performed on the same bench with a Cs-Ne cell is reported for reference since Neon permeation is supposed to be much lower. Windows with missing data correspond to breaks or crashes of the clock software. The clock frequency at $t =$~0 is 9192 736 114 Hz and 9 192 715 170 Hz, for \ac{ASG} and \ac{BSG} cells, respectively. (b)~Evolution of the buffer gas pressure in the cell derived (Eq.~(\ref{eq:eq3})) from frequency measurements shown in (a), where extrapolated data follow an exponential model based on Eq.~(\ref{eq:eq1}).
}
\label{fig:fig2}
\end{figure}

Figure \ref{fig:fig2}(a) shows typical temporal traces of the clock frequency, for two different Cs-He microfabricated cells, one with \ac{BSG} and one with \ac{ASG}, resulting from data acquisitions longer than 40 days. In here, the reduction of the clock frequency with time is compliant with He atoms exiting progressively the cell. In that sense, this experiment is different than the one reported in Ref.~\citenum{Dellis:2016} where a positive slope was observed, due to He entering into the cell placed in a pressurized chamber under He atmosphere. For comparison, a similar measurement was performed on a Cs microcell with \ac{BSG} windows and filled with about 70~Torr of Ne. Since Ne permeation rate is known to be more than three orders of magnitude lower than He~\cite{Abdullah:APL:2015}, the \ac{BSG}-Ne cell is used here as a reference and sets the measurement background of our setup. Focusing onto the first 10 days, linear fitting of clock frequency data yields frequency variations of $-1182$, $-3.5$ and $-0.3$ Hz/day for the \ac{BSG}, \ac{ASG} and \ac{BSG}-Ne reference cell, respectively. These results show that \ac{ASG} drastically reduces He permeation, leading here during the first days to a clock frequency drift reduction by a factor of nearly $350$ compared to \ac{BSG}-based cells.\\
Following such a measurement, the clock frequency is converted into buffer gas pressure using Eq.~(\ref{eq:eq3}). The corresponding evolution of the buffer gas pressure is shown in Fig.~\ref{fig:fig2}(b). Pressure data are then fitted with Eq.~(\ref{eq:eq1}) to estimate the time constant $\tau$, whereas the permeation constant $K$ is extracted using Eq.~(\ref{eq:eq2}). In such measurements of several tens of days, the exponential behavior of the buffer gas pressure evolution is clearly visible for the \ac{BSG}-based cells. The corresponding time constant $\tau$ is $75\pm6$~days, whereas it is extended to $35000\pm3000$~days in \ac{ASG}-based cells. Hence, derived permeation constant $K$ for \ac{BSG} and \ac{ASG}, are $K$~=~5.9$\pm$0.7~$\times$~10$^{-19}$, and 1.3$\pm$0.2~$\times$~10$^{-21}$ m$^{2}$.s$^{-1}$.Pa$^{-1}$, respectively. Concerning \ac{ASG}, our measurements are in good agreement with the ones of Dellis~\textit{et al.}~\cite{Dellis:2016} despite a different cell temperature and confirm its significant capabilities for reduction of He permeation.
In addition, note that permeation constant for BSG-Ne cells is found to be $K$~=~3.5~$\pm$0.2$\times$~10$^{-22}$ at 70$^{\circ}$C, close to the value of $K$~=~5.7$\pm$0.7~$\times$~10$^{-22}$ measured at a higher temperature of 81$^{\circ}$C by Abdullah~\textit{et al.}~\cite{Abdullah:APL:2015} 

\begin{figure*}[t]
\centering
\includegraphics[width=0.90\linewidth]{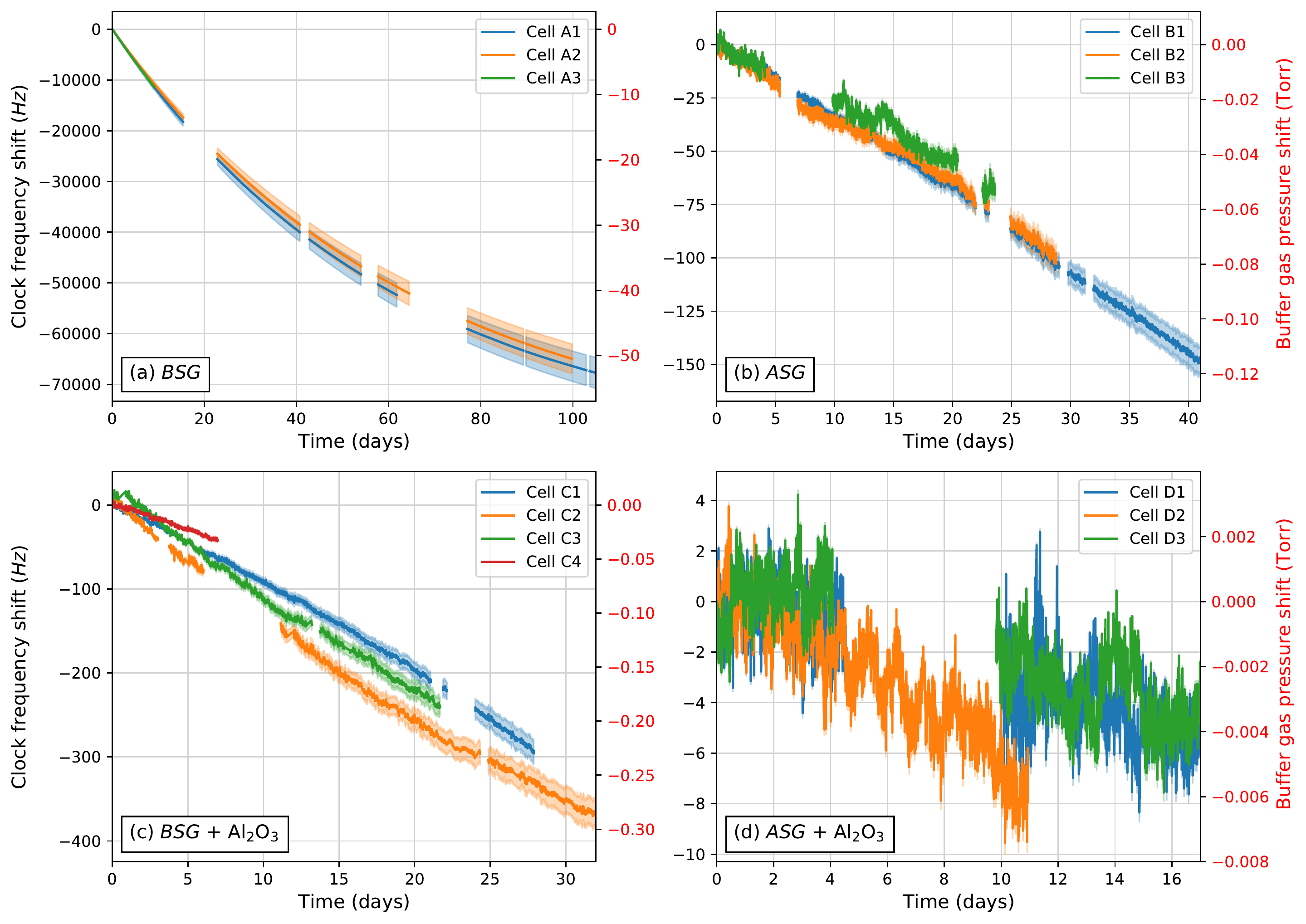}
\caption{(Color online) Temporal trace of the clock frequency for the 4 different configurations of Cs-He cells based on: (a) \ac{BSG}, (b) \ac{ASG}, (c) \ac{BSG} coated with Al$_2$O$_3$ (d) \ac{ASG} coated with Al$_2$O$_3$. The right vertical axis converts the clock frequency evolution into the actual buffer gas pressure decrease. For clarity and easier comparison between all the cells, the initial pressure value was offset to 0 at $t=$~0. Actual initial buffer gas pressure values are given in the text. Light-colored zones indicate the size of error bars attributed to the buffer gas pressure estimation.}
\label{fig:figure3}
\end{figure*}

In this study, we propose to investigate the further permeation reduction obtained with additional thin-film coatings. Al$_2$O$_3$ layers, for instance, have been known and employed to avoid alkali consumption in vapor cells, notably for Cs~\cite{Woetzel:2015}, Rb~\cite{Karlen:2017}, or more recently Sr~\cite{Pate:2023}. The deposition of Al$_2$O$_3$ layers used in our experiments was performed by \ac{ALD} with a commercial Beneq TFS-200 system. 20~nm-thick films have been coated directly on the glass wafers prior to the bonding with silicon and thus cover the inner side of the glass windows. Among the eight cell-wafers developed for this study, five were coated with Al$_2$O$_3$. 
Figure~\ref{fig:figure3}(a)-(d) reports clock frequency measurements performed on several typical cells for each of the four tested configurations, i.e. \ac{BSG}, \ac{ASG} as well as \ac{BSG} and \ac{ASG} coated with Al$_2$O$_3$. All these clock measurements were performed at a cell temperature of nearly 70$^{\circ}$C. As shown, the frequency variations strongly differ from one configuration to another. Standard \ac{BSG} cells (Fig.~\ref{fig:figure3}(a)), characterized by an initial He pressure of $72.5\pm0.5$~Torr exhibit a rapid decrease of the clock frequency, measured to be about $-$1200~Hz/day during the first 15~days. We then measure that about 50~Torr of He escaped the cell after 100~days. These results yield an estimated permeation constant $K$ in the range of 5.5 to 6.3$\pm$0.7~$\times$~10$^{-19}$~m$^2$.s$^{-1}$.Pa$^{-1}$. 

Cells with \ac{ASG} (Fig.~\ref{fig:figure3}(b)) featuring an initial He pressure of $84\pm2$~Torr lead to a drastic reduction of the clock frequency change with a slope of about $-$3.4~Hz/day. This corresponds to a He pressure loss of about $120$~mTorr after 40~days. In comparison with \ac{BSG}, this corresponds to a reduction of the permeation constant $K$ by about 450, yielding $K$ in the range of 1.09 to 1.43$\pm$0.14~$\times$~10$^{-21}$ m$^2$.s$^{-1}$.Pa$^{-1}$ and a time constant increased from 75 to 35000~days. This improvement is in the same order than the one observed in Ref.~\citenum{Dellis:2016}.

Figure~\ref{fig:figure3}(c) displays measurements on \ac{BSG} cells coated with Al$_2$O$_3$. Reported frequency shifts were measured on 4 different cells taken out of three different wafers. A slight dispersion is experienced, resulting first from different initial pressures, and that may be also attributed to a lack of homogeneity in the deposition of Al$_2$O$_3$ layers or their partial damaging during subsequent cell fabrication or dispenser activation. Nevertheless, the shift rate is significantly lower, i.e. nearly 100$\times$ lower than for \ac{BSG} cells, which corresponds to an improvement of the permeation constant of 110, leading to values ranging from 4.6 to 6.3$\pm$0.5~$\times$~10$^{-21}$ m$^2$.s$^{-1}$.Pa$^{-1}$. Note that such permeation is only $4\times$ larger than the one of \ac{ASG} and demonstrates that Al$_2$O$_3$ contributes to reduce He permeation through the cell glass windows. Finally, Fig.~\ref{fig:figure3}(d) shows the clock frequency as a function of time for \ac{ASG}-based cells coated with Al$_2$O$_3$. For e.g. D3 cell, whose pressure is initially measured at $39.6\pm0.4$~Torr, the clock frequency shift is reduced by a factor of 10 ($-$0.3~Hz/day), leading to a loss of 4~mTorr of He after more than two weeks at 70$^{\circ}$C. The corresponding permeation constant is equal to 2.2$\pm$0.2~$\times$~10$^{-22}$ m$^2$.s$^{-1}$.Pa$^{-1}$, i.e. 5.8$\times$ smaller than for uncoated \ac{ASG}. It can be underlined that such permeation value for He is even better than the one for Ne in \ac{BSG} and allow us to envision using He as a buffer gas in microfabricated cells.

\begin{figure*}[t]
\centering
\includegraphics[width=0.95\linewidth]{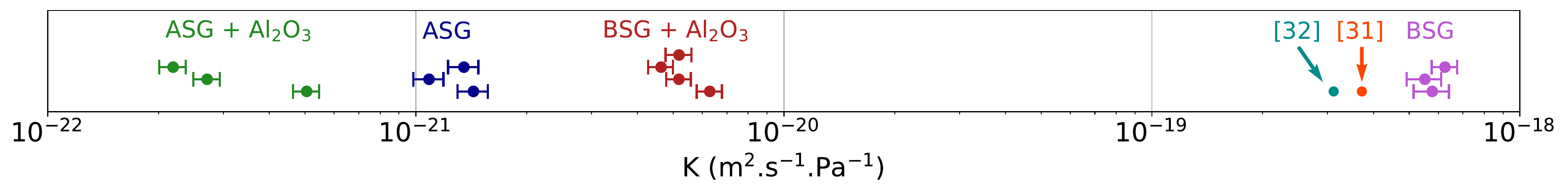}
\caption{(Color online) Distribution of He permeation constants $K$ measured for the 13 tested cells, split in 4 different windows configurations.}
\label{fig:figure4}
\end{figure*}

Figure~\ref{fig:figure4} summarizes of the permeation constant $K$ derived for all of the 13 measured cells. Some values extracted from the literature~\cite{Norton:1957, Altemose:1961} for \ac{BSG} are also indicated. Note that all the above measurements have been performed at nearly 70$^{\circ}$C. Nevertheless, in order to predict and eventually simulate the behavior of microfabricated cells with time and at different temperatures, it is important to determine the activation energy $E_D$ corresponding to a specific glass substrate. If, in the literature, numerous data can be found on \ac{BSG}, and in particular on Pyrex, including $E_D$ (equal to, e.g. $-$0.29~eV~\cite{Rogers:1954, Norton:1957, Altemose:1961}), it seems that no measurement of SD2 \ac{ASG} has been reported yet. In Ref.~\citenum{Dellis:2016}, $E_D$ of SD2 \ac{ASG} was assumed to be comparable to the one measured for another \ac{ASG}, the Corning 1720 glass having close chemical composition (but lower amount of Al$_2$O$_3$, i.e. 18$\%$ instead of 25$\%$) and for which the corresponding $E_D$ was measured equal to $-$0.52~eV in Refs.~\citenum{Norton:1957} and \citenum{Altemose:1961}.

\begin{figure}[t]
\centering
\includegraphics[width=0.99\linewidth]{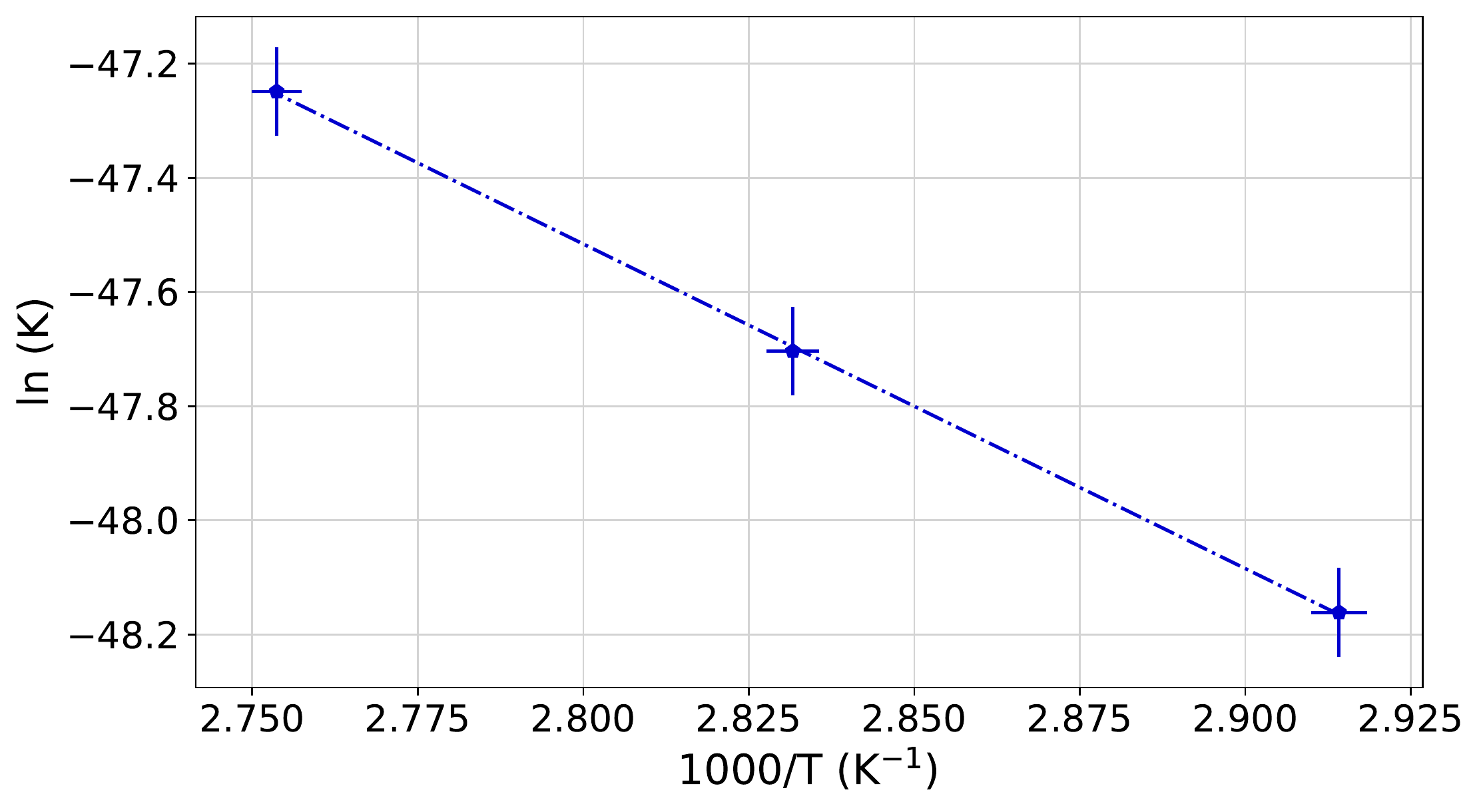}
\caption{Permeation constant $K$ (logarithm) versus the inverse of the cell temperature $T$ (1000/$T$), for an \ac{ASG}-based cell.}
\label{fig:figure5}
\end{figure}

In this study, we aim at estimating the value of $E_D$ for SD2 \ac{ASG} by measuring the clock frequency evolution of the B2 cell at 3 different temperatures. The energy of activation $E_D$ can then be extracted through linearization of the dependence $\ln (K) = \ln (K_0) - \frac{E_D}{k_B T}$, with $T$ the cell temperature and $k_B$ the Boltzmann constant. Figure~\ref{fig:figure5} shows the corresponding result. The fit of experimental data yields $-E_D/k_B$~=~5680 K and $K_0$~=~1$\times$10$^{-14}$~m$^2$.s$^{-1}$.Pa$^{-1}$ from which is derived the activation energy $E_D~\simeq$~$-0.49$~eV. This value is slightly smaller than the one measured for the Corning 1720 glass ($-$0.52~eV).

According to this result, we can, as perspectives, estimate the lifetime of \ac{ASG}-based microfabricated cells in two examples of applications: the first one concerns an atomic clock where He would be employed in a gas mixture along with Ne to shift the cell operational temperature toward higher values~\cite{Kroemer:OE:2015}. In this case, the relatively small required amount of He (e.g. 4.5$\%$ of the mixture for 100$^{\circ}$C, thus around 3~Torr) is favorable so that the inversion temperature should not change by more than $1^{\circ}$C for nearly 500 days. The second application could be cells dedicated to host magneto-optical traps placed in earth atmosphere at room temperature.  For such applications, the inner pressure should remain below 1~$\times$~10$^{-6}$~mbar~\cite{Burrow:2021}, value which would be achieved after 85 days if only influenced by He permeation. Note that such estimated lifetime should be significantly extended with Al$_2$O$_3$ coatings, and reach almost 500 days provided that a similar enhancement ($\times$5.8 at 70$^{\circ}$C) is observed at room temperature. 

In conclusion, we have investigated the permeation of helium through the glass windows of microfabricated alkali vapor cells by measuring the frequency shifts encountered by a \ac{CPT}-based atomic clock. A total of 13 cells have been tested, divided in four configurations, either relying on \ac{BSG} or \ac{ASG} windows, or on \ac{BSG} or \ac{ASG} coated with Al$_2$O$_3$ layers. We have demonstrated that Al$_2$O$_3$ coatings, usually employed to prevent the consumption of alkali in vapor cells, can also significantly reduce He permeation. Cells using \ac{BSG} coated with Al$_2$O$_3$ feature permeation constants about 110 times smaller than those based on \ac{BSG} only. These performances are only 4 times lower than the ones measured for \ac{ASG} for which permeation constants of about 1.1~$\times$~10$^{-21}$~m$^2$.s$^{-1}$.Pa$^{-1}$ were found, at a temperature of 70$^{\circ}$C. Al$_2$O$_3$ coatings seem to be also efficient when deposited on \ac{ASG}, further improving the hermeticity of cells with permeation constant reaching about 2.2~$\times$~10$^{-22}$~m$^2$.s$^{-1}$.Pa$^{-1}$, i.e. almost twice less than for \ac{BSG}-Ne cells employed in miniature atomic clocks. Finally, the activation energy for \ac{ASG} was estimated to be about $-$0.49~eV from measurements of the permeation constant at three different cell temperatures.

\section*{Acknowledgments}
This work was supported partially by the Direction G\'{e}n\'{e}rale de l’Armement (DGA) and by the Agence Nationale de la Recherche (ANR) in the frame of the ASTRID project named PULSACION under Grant ANR-19-ASTR-0013-01 as well as by the R\'{e}gion Franche-Comt\'{e} (NOUGECELL project). It was also supported by the Agence Nationale de la Recherche (ANR) in the frame of the LabeX FIRST-TF (Grant ANR 10-LABX-48-01), the EquipX Oscillator-IMP (Grant ANR 11-EQPX-0033) and the EIPHI Graduate school (Grant ANR-17-EURE-0002). The work of Cl\'{e}ment Carl\'{e} was funded by the Centre National d'Etudes Spatiales (CNES) and the Agence Innovation D\'{e}fense (AID) and the one of Petri Karvinen by the Photonics Research and Innovation (PREIN) Programme of the Academy of Finland, decision 320166. This work was partly supported by the french RENATECH network and its FEMTO-ST technological facility (MIMENTO).

\section*{Author Declarations}
\subsection*{Conflict of Interest}
The authors have no conflicts to disclose.
\subsection*{Author Contributions}
\textbf{Cl\'ement Carl\'e:} Data Curation (lead); Formal Analysis (lead); Methodology (lead); Software (lead); Visualization (lead); Writing/Original Draft Preparation (equal). Writing/Review $\&$ Editing (equal). \textbf{Shervin Keshavarzi:} Methodology (equal); Resources (equal). \textbf{Andrei Mursa:} Resources (equal). \textbf{Petri Karvinen:} Resources (equal). \textbf{Ravinder Chutani:} Resources (equal). \textbf{Sylwester Bargiel:} Resources (equal). \textbf{Samuel Queste:} Resources (equal). \textbf{R\'emy Vicarini:} Data Curation (equal); Resources (equal). \textbf{Philippe Abb\'e:} Resources (equal). \textbf{Moustafa Abdel Hafiz:} Data Curation (equal); visualisation (supporting); Writing/Review $\&$ Editing (supporting). \textbf{Vincent Maurice:} Resources (equal); Software (equal); Writing/Review $\&$ Editing (supporting). \textbf{Rodolphe Boudot:} Conceptualization (equal); Data Curation (supporting); Formal Analysis (supporting); Funding Acquisition (lead); Methodology (supporting); Project Administration (lead); Visualization (equal); Writing/Original Draft Preparation (equal); Writing/Review $\&$ Editing (supporting). \textbf{Nicolas Passilly:} Conceptualization (lead); Data Curation (equal); Formal Analysis (supporting); Funding Acquisition (equal); Methodology (equal); Project Administration (equal); Resources (lead); Visualization (equal); Writing/Original Draft Preparation (lead); Writing/Review $\&$ Editing (lead).

\section*{Data availability statement}
The data supporting the findings of this study are available from the corresponding author upon reasonable request.

\section*{References}
\bibliography{CARLE_bib}
\end{document}